\documentclass[aps,showpacs,preprint,floats]{revtex4}

\usepackage{graphicx}

\begin{document}

\begin{flushright}
\vspace*{.4in}
TUM-HEP-463/02\\
MADPH-01-1270
\end{flushright}

\title{Neutrinos Associated With Cosmic Rays of Top-Down Origin}
\author{C. Barbot$^1$, M. Drees$^1$, F. Halzen$^2$, D. Hooper$^2$} 
\address{$^1$Physik Dept., TU M\"unchen, James Franck Str., D-85748
Garching, Germany\\   
$^2$Department of Physics, University of Wisconsin, 1150 University Avenue,   
Madison, WI, USA 53706}
\date{\today} 

\begin{abstract}
Top-down models of cosmic rays produce more neutrinos than photons and
more photons than protons.  In these models, we reevaluate the fluxes
of neutrinos associated with the highest energy cosmic rays in light
of mounting evidence that they are protons and not gamma rays. While
proton dominance at EeV energies can possibly be achieved by efficient
absorption of the dominant high-energy photon flux on universal and
galactic photon and magnetic background fields, we show that the
associated neutrino flux is inevitably increased to a level where it
should be within reach of operating experiments such as AMANDA II,
RICE and AGASA. In future neutrino telescopes, tens to a hundred,
rather than a few neutrinos per kilometer squared per year, may be
detected above 1 PeV.
\end{abstract}
\pacs{98.70.5a, 95.35.+d, 95.85.Ry}

\maketitle
\thispagestyle{empty}

\section{Introduction}

The discovery of cosmic rays with energy exceeding the GZK cutoff
\cite{gzk} presents an interesting challenge to astrophysics, particle
physics, or both \cite{agasa,hires}. Numerous scenarios have been
proposed to solve the problem. These include exotic particles
\cite{exo}, neutrinos with QCD scale cross sections \cite{qcdneu},
semi-local astrophysical sources \cite{loc} and top-down models
\cite{top}.

Top-down models can be motivated by a variety of arguments. For
example, the recent measurements of the cosmic microwave background
and of supernova redshifts have dramatically confirmed that our
universe contains a large fraction of cold dark matter \cite{cmb}.  A
top-down model in which annihilating or decaying superheavy relic
particles produce the highest energy cosmic rays could potentially
solve both of these problems \cite{wim1,wim}. Topological defects
could also solve the ultra-high energy cosmic ray problem in a similar
way \cite{top}.  In this paper, we reevaluate the implications of
generic top-down models producing parton jets of $10^{21}$ or
$10^{25}$ eV that fragment into the observed super GZK cosmic rays.

Established particle physics implies that such ultra high-energy jets
fragment predominantly into pions and kaons, with a small admixture of
protons \cite{lep}. The mesons will eventually decay into photons or
electrons plus neutrinos. A typical QCD jet therefore produces more
photons than protons. This is true in particular at relatively low
values of $x = E_{\rm particle} / E_{\rm jet}$, but even at large $x$
the photon flux is at least as large as the proton flux in a jet.
This seems to be in disagreement with mounting evidence that the
highest energy cosmic rays are not photons \cite{pro1}. The observed
shower profile of the original Fly's Eye event, with energy exceeding
$10^{20}$ eV, fits the assumption of a primary proton, or, possibly,
that of a nucleus.  The shower profile information is sufficient to
conclude that the event is unlikely to be of photon origin.  The same
conclusion is reached for the Yakutsk event that is characterized by a
large number of secondary muons, inconsistent with a purely
electromagnetic cascade initiated by a gamma ray.  A recent reanalysis
of Haverah Park data further reinforces this conclusion \cite{park}.
In light of this information, it seems likely that protons, and not
gamma rays, dominate the highest energy cosmic ray spectrum. This does
not necessarily rule out superheavy particles as the source of the
highest energy cosmic rays.  The uncertainties associated with the
cascading of the jets in the universal and galactic radio backgrounds
and with the strength of intergalactic magnetic fields leave open the
possibility that ultra high-energy photons may be depleted from the
cosmic ray spectrum near $10^{20}$ eV, leaving a dominant proton
component at GZK energies \cite{lee,radio}.  With this in mind, we
will choose to normalize the proton spectrum from top-down scenarios
to the observed ultra high-energy cosmic ray flux.

Neutrinos are produced more numerously than protons and travel much
greater distances. The main point of this paper is to point out that
this ``renormalization'' of the observed cosmic ray flux to protons
generically predicts observable neutrino signals in operating
experiments such as AMANDA II, RICE and AGASA. Top-down models, if
not revealed, will be severely constrained by high-energy neutrino
observations in the near future.

\section{Nucleons from Ultra-high energy Jets}

The assumption that nucleons from the decay (or annihilation) of very
massive $X$ particles are the source of the highest energy cosmic rays
normalizes the decay or annihilation rate of their sources, once the
shape of the spectrum of the produced nucleons is known. One needs
mass $M_X \geq 10^{21}$ eV in order to explain the observed UHECR
events. The presence of such very massive particles strongly
indicates the existence of superparticles with masses at or below the
TeV scale, since otherwise it would be difficult to keep the weak
energy scale ten or more orders of magnitude below $M_X$ in the
presence of quantum corrections. Moreover, we know that all gauge
interactions are of comparable strength at energies near $M_X$. These
two facts together imply that the evolution of a jet with energy $\geq
10^{21}$ eV shows some new features not present in jets produced at
current particle collider experiments.

First of all, primary $X$ decays are likely to produce approximately
equal numbers of particles and superparticles, since $M_X$ is much
larger than the scale $M_{\rm SUSY} \leq 1$ TeV of typical
superparticle masses. Even if the primary $X$ decay only produces
ordinary particles, superparticles will be produced in the subsequent
shower evolution. Note also that (at least at high energies)
electroweak interactions should be included when modeling the parton
shower. Both effects taken together imply that the jet will include
many massive particles -- superparticles, electroweak gauge and Higgs
bosons, and also top quarks. The decays of these massive particles
increase the overall particle multiplicity of the jet, and also
produce quite energetic neutrinos, charged leptons and lightest
supersymmetric particles (LSP). Eventually the quarks and gluons in
the jet will hadronize into baryons and mesons, many of which will in
turn decay.

We model these jets at the point of their origin using the program
described in Ref.\cite{bd1}. This program allows us to calculate
spectra for different $X$ decay modes. It then follows the
supersymmetric parton cascade down to virtuality (or inverse time) of
the order of $M_{\rm SUSY}$, including all gauge interactions as well
as third generation Yukawa interactions. At $M_{\rm SUSY}$ all massive
particles are decoupled from the parton shower, and decay.
Supersymmetric cascade decays are fully taken into account; the
results presented below have been obtained using the same spectrum of
superparticles as in ref.\cite{bd1}. At virtualities below $M_{\rm
SUSY}$ only ordinary QCD interactions contribute significantly to the
development at the jet; $b$ and $c$ quarks are decoupled at their
respective masses, hadronize, and decay. At a virtuality near $1$ GeV
the light quarks and gluons hadronize, with a meson to baryon ratio of
roughly thirty to one (five to one) at small (large) $x$. All baryons
will eventually decay into protons, while the mesons (mostly pions)
decay into photons, electrons \cite{footnote} and neutrinos (plus their
antiparticles). The heavier charged leptons (muons and taus) also
decay. The final output of the code is the spectra of seven types of
particles which are sufficiently long--lived to reach the Earth:
protons, electrons, photons, three flavors of neutrinos, and LSPs. We
assume that $X$ decays are CP--symmetric, i.e. we assume equal fluxes
of particles and antiparticles of a given species.

The calculation of Ref.\cite{bd1} was based on conventional one--loop
evolution equations for the relevant fragmentation functions. These
may not be reliable in the region of very small $x$. We wish to
calculate neutrino fluxes at energies down to $\sim 10^{15}$ eV (1
PeV), which corresponds to $x \sim 10^{-6} \ (10^{-10})$ for $M_X =
10^{21} \ (10^{25})$ eV. At these very small $x$ values color
coherence effects are expected to suppress the shower evolution
\cite{mlla}. We try to estimate the size of these effects by matching
our spectra computed using conventional evolution equations to the
so--called asymptotic MLLA spectra; details of this procedure will be
described elsewhere \cite{bd2}. The effect of this modification on the
neutrino event rate is relatively modest for primary jet energy near
$10^{21}$ eV, but becomes significant at $10^{25}$ eV. However, even
at this higher energy the proton flux, which we only need at $x \geq
10^{-5}$, is not affected significantly.

\begin{figure}[thb]
\centering\leavevmode
\includegraphics[width=5in]{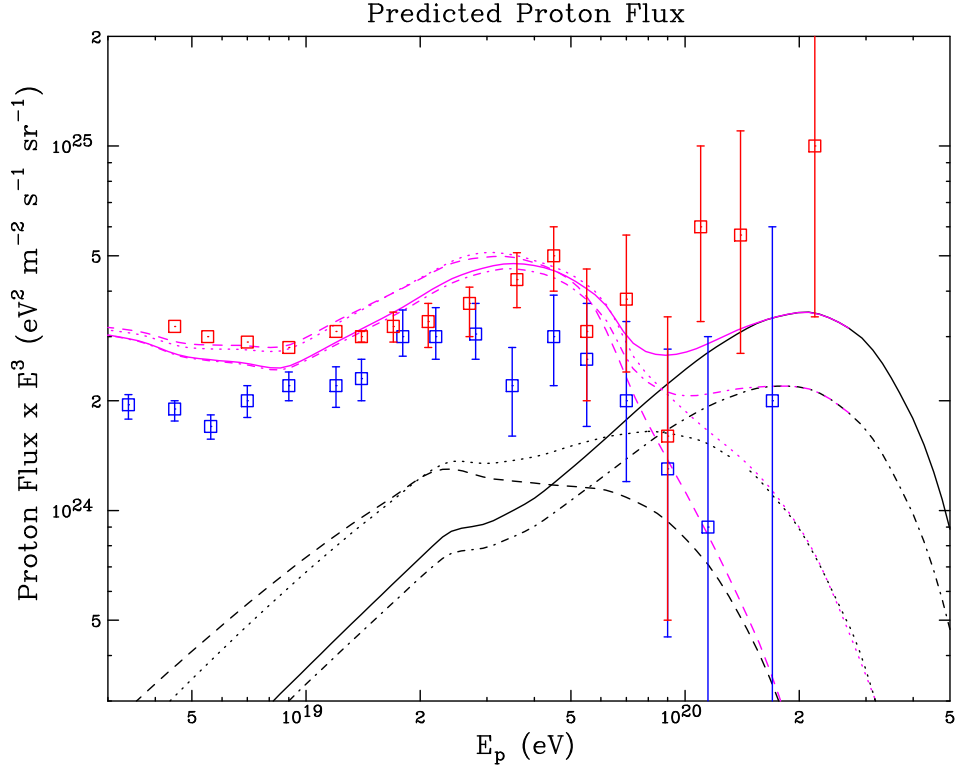}
\caption{The ultra high-energy cosmic ray flux predicted for the decay
of superheavy particles with mass $M_X = 2 \cdot 10^{21}$ eV is
compared to the HIRES (darker) and AGASA (lighter) cosmic ray data.
The distribution of jets used includes an overdensity
factor of $10^5$ within 20 kpc of the galaxy. Spectra are shown for
quark$+$antiquark (solid), quark$+$squark (dot-dash), $SU(2)$ doublet
lepton$+$slepton 
(dots) and 5 quark$+$5 squark (dashes) initial states.  Dark lines are
from top-down origin alone whereas lighter lines are top-down plus an
homogeneous extragalactic contribution as predicted in Ref.\cite{lee}. Note that all observed super GZK events can be explained
by this mechanism.}
\label{one}
\end{figure}

\begin{figure}[thb]
\centering\leavevmode
\includegraphics[width=5in]{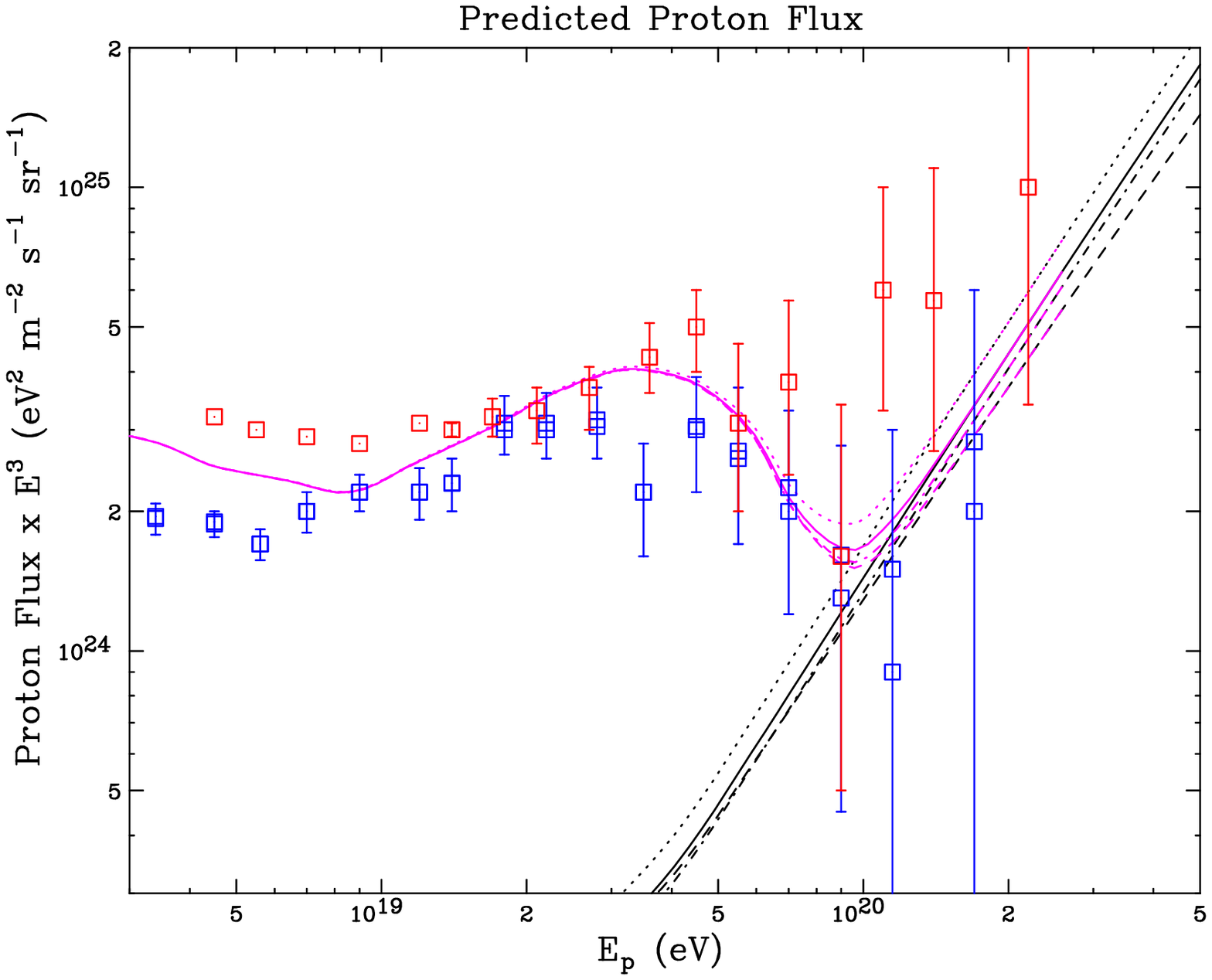}
\caption{As in figure 1, but using particles of mass $M_X = 2 \cdot
10^{25}$ eV.} 
\label{two}
\end{figure}

This calculation gives us the shape of the spectra of the stable
particles at source. The spectra on Earth might differ significantly
due to propagation effects. As stated in the introduction, we will
assume that (almost) all UHE photons get absorbed. This is actually
expected to be true for a homogeneous source distribution. However,
according to current estimates of the strengths of the magnetic fields
and of the radio wave background in (the halo of) our own galaxy most
UHE photons produced in the halo of our galaxy are expected to reach
the Earth. As stated in the Introduction, this seems to be in conflict
with observation. We will therefore assume that the interaction length
of UHE photons in our galaxy has been greatly over--estimated, and
explore the consequences of this assumption for neutrino signals.

As well known, (anti)protons lose energy when traveling through the
intergalactic medium, mostly through scattering off photons of the
ubiquitous cosmic microwave background (CMB). We calculate the
observed spectrum of protons taking into account scattering off the
CMB at the $\Delta-$resonance and scattering by $e^+e^-$ pair
production; energy losses through the Hubble expansion of the Universe
are also included \cite{lee,prop}. Note that the photoproduction of
charged pions contributes to the observed neutrino flux on Earth. In
order to solve the ultra high-energy cosmic ray problem, the
(anti)proton flux must accommodate the events above the GZK cutoff.
Observations indicate on the order of a few times $10^{-27}$ events
$\rm{m}^{-2} \rm{s}^{-1} \rm{sr}^{-1} \rm{GeV}^{-1}$ in the energy
range above the GZK cutoff ($5 \times 10^{19}$ eV to $2 \times
10^{20}$ eV)\cite{hires,agasa}.

The formalism of a generic top-down scenario is sufficiently flexible
to explain the data from either the HIRES \cite{hires} or AGASA
\cite{agasa} experiments.  Figure 1 compares HIRES and AGASA data to
the proton spectrum predicted for a galactic distribution of decaying
particles with mass $M_X = 2 \cdot 10^{21}\,$ eV. The drop near a few times $10^{19}$ eV is a manifestation of the GZK cutoff. Note,
however, that there are sufficient semi-local events to explain all
observed super GZK events.  Similarly, figure 2 compares HIRES and
AGASA data to the spectrum predicted for $M_X = 2 \cdot 10^{25}$ eV,
rather than $ 2 \cdot 10^{21}$ eV, decaying particles for the same
distribution. Although HIRES and AGASA data differ at face value,
especially above the GZK cutoff, top-down scenarios can accommodate
all events observed above the GZK cutoff in either experiment.

If the cosmic ray sources are not distributed with a large overdensity
in the galaxy, the resulting cosmic ray and neutrino spectrum will be
modified.  For example, using a homogeneous distribution, the GZK
cutoff will again be manifest and the observed cosmic ray spectrum
will be difficult to explain. A galactic overdensity of $10^3$ to
$10^4$ or more seems necessary to fit the data. The figure 1 shows a
$10^5$ overdensity, which is the overall overdensity of matter in our
galaxy at the location of the Sun. Note that for less extreme
overdensities, the average distance at which a proton is produced will
be larger. This implies larger energy losses, and hence a reduced
proton flux on Earth for a given number of sources.  Conversely, if we
fix the proton flux to the observed flux of UHECR events, models with
lower overdensity require more sources. Since neutrino fluxes are not
degraded by propagation through the intergalactic medium, the number
of neutrinos increases proportionally to the number of sources, with
additional contributions to the neutrino flux coming from pion
production on the CMB background.  Thus, the neutrino event rates and
spectrum shown in the figures reflect the most conservative choice of
distributions.  The table shows results for both homogeneous and galactic distributions.

\section{Neutrinos from Ultra high-energy Jets}

As discussed earlier, the program computing the proton flux at source
also gives the neutrino flux at source.  Neutrinos, not being limited
by scattering, travel up to the age of the universe at the speed of
light ($\sim$ 3000 Mpc in an Euclidean approximation). The only
nontrivial effect of neutrino propagation is due to oscillations. In
our case the propagation distance of neutrinos amounts to many
oscillation lengths, if oscillation parameters are fixed by the
currently most plausible solutions of the atmospheric and solar
neutrino deficits \cite{bahcall}. As a result, the UHE neutrino flux
on Earth is the same for all three flavors, and amounts to the average
of the fluxes of the three neutrinos flavors at source. 

The predicted neutrino flux is shown in figures 3 and 4. At $E_\nu \ll
E_{\rm jet}$ the main contribution comes from $\pi^\pm \rightarrow
\mu^\pm \nu_\mu \rightarrow e^\pm \nu_e \nu_\mu$ decays, but at larger
$E_\nu$ there can be significant contributions from the decays of
heavy (s)particles. The peak in the dotted curves at $E_\nu = E_{\rm
jet}$ results from our assumption that in this scenario $X$ decays
directly into first or second generation $SU(2)$ doublet (s)leptons,
which implies that 50\% of all $X$ decays give rise to a primary
neutrino; in this case the ratio of neutrino and proton fluxes has a
maximum at high energy. On the other hand, if primary $X$ decays are
purely hadronic, the neutrino flux at the largest energy is only
slightly above the proton flux at that energy. The reason is that
neutrinos from meson decays only carry a fraction of the energy of the
meson, so a five to one meson to proton ratio at large $x$ leads to a
nearly one to one neutrino to proton ratio. We see that the neutrino
flux at the highest energy depends quite strongly on how the $X$
particles decay; there is also some dependence on the parameters of
the SUSY model \cite{bd1,bd2}. For given proton flux the neutrino flux
at smaller $x$ is much less model dependent. At very small $x$ a new
uncertainty appears due to coherence effects. These have so far only
been studied in a pure QCD parton shower; our treatment of these
effects is therefore of necessity rather crude.

\begin{figure}[thb]
\centering\leavevmode
\includegraphics[width=5in]{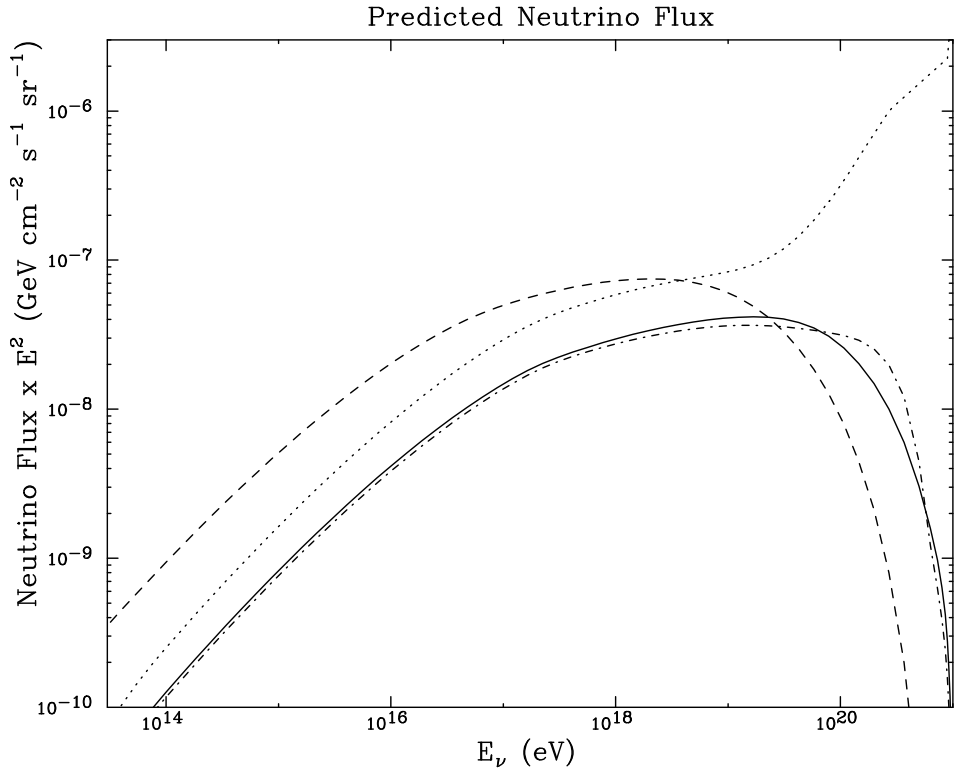}
\caption{The neutrino plus anti-neutrino flux corresponding to the
cosmic ray spectra of 
figure 1 from the decay of superheavy particles with mass $M_X = 2
\cdot 10^{21}$ eV.
Spectra are shown for quark-antiquark (solid),
quark-squark (dot-dash), lepton-slepton (dots) and 5 quark-5 squark
(dashes) initial states.}
\label{three}
\end{figure}

\begin{figure}[thb]
\centering\leavevmode
\includegraphics[width=5in]{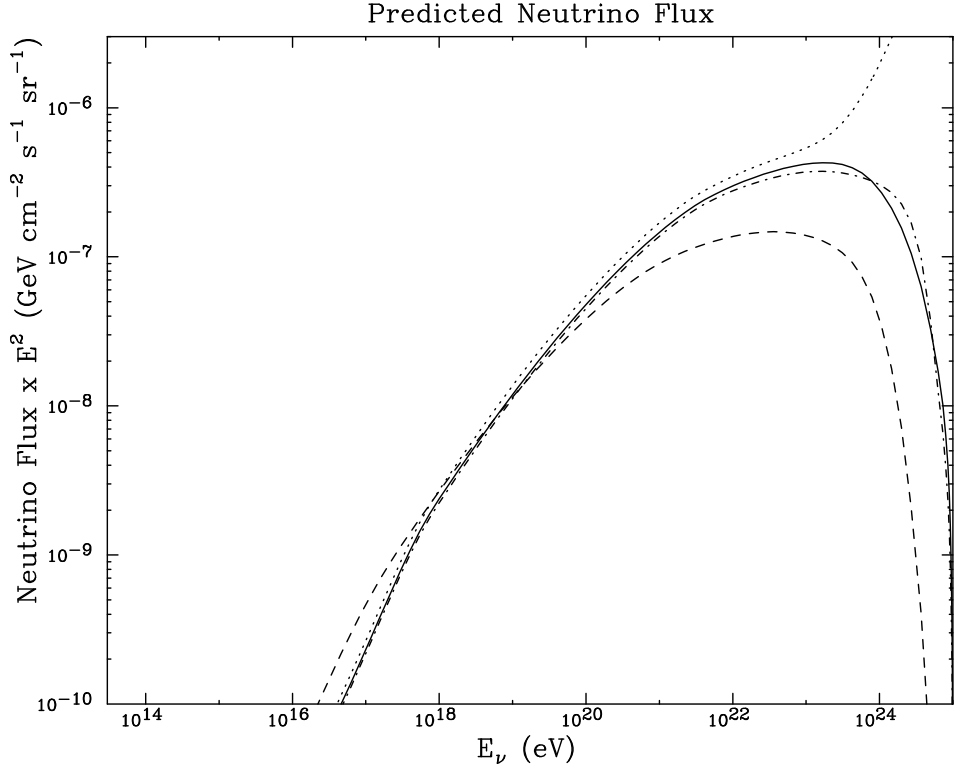}
\caption{As figure 3, but corresponding to the cosmic ray spectrum of
figure 2 with $M_X = 2 \cdot 10^{25}$ eV.} 
\label{four}
\end{figure}

\section{Event Rates in High-energy Neutrino Telescopes and Air Shower
Experiments} 

We will discuss two classes of experiments capable of observing high
energy cosmic neutrinos: neutrino telescopes and air shower
experiments.

Optical Cerenkov neutrino telescopes such as the operating AMANDA II
and next generation IceCube are designed to observe muon tracks from
charged current interactions as well as showers which occur in the detector.
The probability of detecting a neutrino passing through the detector
from its muon track is given by
\begin{equation}
P_{\nu \rightarrow \mu}(E_\nu, \theta_{\rm zenith}) = \sigma_{\nu
N}(E_\nu)\, n_{\rm{H_2 O}} \, R_\mu(E_\mu, \theta_{\rm zenith}) 
\end{equation}
where $n_{\rm{H_2 O}}$ is the number density of nucleons in the
detector medium (water or ice), and the muon range
$R_{\mu}(E_{\mu},\theta_{\rm{zenith}})$ is the average distance
traveled by a muon of energy $E_{\mu}$ before falling below some
threshold energy (we have used 100 TeV).  This quantity depends on the
zenith angle of the incoming neutrino because for a detector depth of
$\sim 2\,$km, only quasi-horizontal or upgoing events can benefit from
longer muon ranges.  At the energies we are most concerned with, the
majority of muon events will be quasi-horizontal.
The number of muon events observed is then given by
\begin{equation}
N_{\rm{events}} = \int dE_{\nu}\, d\Omega \,\frac {d \phi_\nu}{dE_\nu}\, P_{\nu
\rightarrow \mu} (E_{\nu},\theta_{\rm zenith})\, A_{\rm eff}\, T, 
\end{equation}
where $T$ is the time observed and $A_{\rm{eff}}$ is the effective
area of the detector: one twentieth square kilometers for AMANDA II
and one square kilometer for IceCube.

AMANDA II and IceCube can also observe showers generated in charged or
neutral current interactions within the detector volume. The event
rate from showers is not enhanced by long muon ranges, but can be
generated by all three flavors of neutrinos and with greater cross
section (neutral + charged current).  We use a shower energy threshold of 100
TeV.  The energy threshold imposed effectively removes any background events from atmospheric neutrino events.  For a review of Optical Cerenkov neutrino telescopes see
Ref.\cite{ice2}.

The operating radio Cerenkov experiment, RICE, is capable of observing
showers generated in charged current electron neutrino events. RICE's
effective volume increases with energy.  At 100 TeV, RICE has an
effective volume less than one hundredth of a cubic kilometer. By 10
PeV, however, it increases to about ten cubic kilometers \cite{rice}.
Again, we use a hard 100 TeV shower threshold.

Air shower experiments can also observe very high energy cosmic
neutrinos.  We consider AGASA, the largest ground array currently in
operation \cite{agasa2}, and the next generation AUGER array
\cite{auger}.

To determine that an air shower was initiated by a neutrino, rather
than a proton or other cosmic ray, we require a slant depth greater
than 4000 $\rm{g}/\rm{cm}^2$. This corresponds to a zenith angle very
near 75 degrees.  Therefore, only quasi-horizontal air shower events
can be identified as neutrinos. Additionally, unlike showers generated in the upper atmosphere,
deeply penetrating showers provide both muon and electromagnetic
shower components which help them be differentiated from showers with
hadronic primaries. The probability of detecting and identifying a neutrino initiated
air shower is described in terms of the array's acceptance, $A$, in
units of volume times water equivalent steradians (we sr).  The
detector's acceptance increases with energy.  For AGASA, the
acceptance is about $0.01 \,\rm{km}^3\,\rm{we}\,{st} $ at $10^7\,$ GeV but
increases to $1.0 \,\rm{km}^3\,\rm{we}\,{st} $ at $10^{10}\,$ GeV and
above.  For AUGER, the acceptance is about $0.1
\,\rm{km}^3\,\rm{we}\,{st} $ at $10^7\,$ GeV, $10.0
\,\rm{km}^3\,\rm{we}\,{st} $ at $10^9\,$ GeV and $50.0
\,\rm{km}^3\,\rm{we}\,{st} $ at $10^{12}\,$ GeV.  The number of events
observed is then
\begin{equation}
N_{\rm events} = \int dE_\nu \, d\Omega \, n_{\rm{H_2 O}} \, \frac{d \phi_\nu} {dE_\nu}
\,\sigma_{\nu N}(E_\nu) \,A(E_\nu) \,T, 
\end{equation}
where $T$ is again the time observed, $n_{\rm{H_2 O}}$ is the number density of nucleons in water and $A(E_{\nu})$ is the
detector's acceptance.  AGASA presently has about five years of
effective running time between 1995 and 2000 analyzed.  A useful
treatment of air shower event rates from neutrinos can be found in
Ref.\,\cite{feng}.

\begin{table}[thb]
\begin{tabular}{c|c c c c c c c c c c c} 
& 
\multicolumn{2}{c }{~AMANDA II~}&  \multicolumn{2}{c }{~AGASA~}  &
\multicolumn{2}{c }{~RICE~~}  & \vline & \multicolumn{2}{c }{~IceCube~}
& \multicolumn{2}{c }{~AUGER~} \\  
\hline \hline
~~$q \bar{q}$, $10^{21}$ eV, Galactic
~~&~~~~0.39~&~~&~~~0.056~&~~&~~11.5~&&~\vline~&~~12.2~&~~&~~~1.5\\
~~$q \tilde{q}$, $10^{21}$ eV, Galactic
~~&~~~~0.36~&~~&~~~0.052~&~~&~~10.7~&&~\vline~&~~11.4~&~~&~~~1.4\\
~~$5 \times q \tilde{q}$, $10^{21}$ eV, Galactic
~~&~~~~1.4~&~~&~~~0.11~&~~&~~33.7~&&~\vline~&~~44.6~&~~&~~~3.1\\
~~$l \tilde{l}$, $10^{21}$ eV, Galactic
~~&~~~~0.96~&~~&~~~0.20~&~~&~~24.5~&&~\vline~&~~29.8~&~~&~~~7.0\\
\hline  \hline 
~~$q \bar{q}$, $10^{25}$ eV, Galactic
~~&~~~~0.041~&~~&~~~0.019~&~~&~~1.1~&&~\vline~&~~1.2~&~~&~~~0.60\\
~~$q \tilde{q}$, $10^{25}$ eV, Galactic
~~&~~~~0.039~&~~&~~~0.018~&~~&~~1.0~&&~\vline~&~~1.1~&~~&~~~0.56\\
~~$5 \times q \tilde{q}$, $10^{25}$ eV, Galactic
~~&~~~~0.039~&~~&~~~0.016~&~~&~~1.1~&&~\vline~&~~1.1~&~~&~~~0.50\\
~~$l \tilde{l}$, $10^{25}$ eV, Galactic
~~&~~~~0.047~&~~&~~~0.022~&~~&~~1.3~&&~\vline~&~~1.4~&~~&~~~0.69\\
~~$q \bar{q}$, no MLLA, $10^{25}$ eV, Galactic
~~&~~~~0.27~&~~&~~~0.041~&~~&~~7.0~&&~\vline~&~~8.9~&~~&~~~1.2\\
\hline  \hline 
~~$q \bar{q}$, $10^{21}$ eV, Homogeneous
~~&~~~~3.5~&~~&~~~0.50~&~~&~~103.8~&&~\vline~&~~110.3~&~~&~~~13.2\\
~~$q \tilde{q}$, $10^{21}$ eV, Homogeneous
~~&~~~~3.2~&~~&~~~0.47~&~~&~~95.9~&&~\vline~&~~102.2~&~~&~~~12.3\\
~~$5 \times q \tilde{q}$, $10^{21}$ eV, Homogeneous
~~&~~~~6.9~&~~&~~~0.57~&~~&~~168.3~&&~\vline~&~~223.2~&~~&~~~15.4\\
~~$l \tilde{l}$, $10^{21}$ eV, Homogeneous
~~&~~~~4.8~&~~&~~~1.0~&~~&~~122.5~&&~\vline~&~~149.2~&~~&~~~35.0\\
\hline  \hline 
~~$q \bar{q}$, $10^{25}$ eV, Homogeneous
~~&~~~~0.62~&~~&~~~0.28~&~~&~~16.5~&&~\vline~&~~18.0~&~~&~9.0\\
~~$q \tilde{q}$, $10^{25}$ eV, Homogeneous
~~&~~~~0.58~&~~&~~~0.27~&~~&~~15.5~&&~\vline~&~~16.9~&~~&~8.5\\
~~$5 \times q \tilde{q}$, $10^{25}$ eV, Homogeneous
~~&~~~~0.58~&~~&~~~0.25~&~~&~~17.0~&&~\vline~&~~17.1~&~~&~7.5\\
~~$l \tilde{l}$, $10^{25}$ eV, Homogeneous
~~&~~~~0.71~&~~&~~~0.33~&~~&~~18.9~&&~\vline~&~~20.7~&~~&~10.3\\
~~$q \bar{q}$, no MLLA, $10^{25}$ eV, Homogeneous
~~&~~~~4.1~&~~&~~~0.61~&~~&~~104.6~&&~\vline~&~~133.2~&~~&~17.9\\
\end{tabular}

\caption{Neutrino events per year in top-down scenarios for several
operating and next generation experiments.  For AMANDA II and IceCube,
100 TeV shower and muon energy thresholds were imposed. Events are
only calculated up to $10^{12}$ GeV as discussed in the text.}
\label{table:I} 
\end{table}

Table 1 shows the event rates expected for a variety of models, and
for several experiments. AMANDA-II, with an effective area of
$\sim$50,000 square meters can place the strongest limits on high
energy neutrino flux presently. Furthermore, AGASA, with five years of
effective observing time, has similar sensitivity.  RICE, just
beginning to release results, will be capable of raising the level to
which top-down scenarios can be tested, perhaps being capable of
testing all 16 models shown in the table. Even if no events are
observed with operating experiments, next generation experiments,
especially IceCube, will be able to test all models with adequate
sensitivity.

Event rates shown in table 1 include only events below $10^{12}$ GeV.
Above this energy, uncertainties in the neutrino-nucleon cross
sections and in detector performance make such calculations difficult
and unreliable.  Our most reasonable extrapolations into this energy
range indicate about a 20\% enhancement to the event rate if all
energies are considered for $10^{25}$ eV jets.  There is no effect for
the $10^{21}$ eV jet case.

High-energy neutrino event rates have been calculated in
Ref.\,\cite{jaime} for a similar model.  Their calculation used the
model of reference \cite{bere} which normalized the ultra high-energy
cosmic ray flux to the photons and protons generated in superheavy
particle decay rather than the proton flux alone. For this reason,
their results show only two events per year in a square kilometer
neutrino telescope, a smaller rate than we predict for most
models. Another recent estimate of neutrino fluxes on Earth in
top--down models \cite{kalashev} finds broadly similar results as
our's. However, there the `MLLA' form for the fragmentation functions
was used for all energies, which (incorrectly) predicts nearly
energy--independent ratios of neutrino, photon and proton fluxes.

\section{Conclusions}

If a top-down scenario, such as the decay or the annihilation of
superheavy relics, is the source of the highest energy cosmic rays, then a
high-energy neutrino flux should accompany the observed cosmic ray
flux.  This neutrino flux will be much higher than the flux of
nucleons due to the much greater mean free path of neutrinos and
greater multiplicity of neutrinos produced in high-energy hadronic
jets.

The high-energy neutrino flux generated in such a scenario can be
calculated by normalizing the flux of appropriate particles to the
ultra--high energy cosmic ray flux. With mounting evidence that the
highest energy cosmic rays are protons or nuclei and not photons, we
have assumed that the ultra high-energy photons are degraded by the
universal and/or galactic radio background, leaving protons to
dominate the highest energy cosmic ray flux. The neutrino flux must
then be normalized to the proton flux resulting in significantly
improved prospects for its detection.

A word about the uncertainties in our calculation might be in
order. First of all, the uncertainty of the measured UHECR flux, and
in particular the discrepancy between the HIRES and AGASA results,
leads to an overall uncertainty of a factor of $2-3$. On the
theoretical side, the main uncertainty probably comes from the
calculation of the particle spectra at ``small'' energies, where
currently not very well understood coherence effects can play a
role. This effect is bigger for higher primary jet energy, and can
change the event rate by up to a factor of about 7 (see table). Relaxing our
assumption that {\em all} UHE photons are absorbed would lead to a
corresponding reduction of the fitted source density, and hence of the
neutrino flux. In this context it is worth mentioning that in the
scenario which seems to fit the data best, with primary jet energy
near $10^{21}$ eV and a galactic source overdensity of about $10^5$ (see
Fig.~1 and ref.\cite{frag2}), including the photon flux fully would
only reduce the predicted event rate by a factor of two to three,
since in this case the flux of $10^{20}$ eV photons at source is only
slightly larger than the corresponding proton flux. This would still
give a neutrino flux in easy striking range of km$^2$ scale detectors.

This paper shows that the neutrino flux accompanying the highest
energy cosmic rays in top-down scenarios is of order of the limits
placed by operating experiments such as AMANDA II, RICE and AGASA.
Further data from these experiments, or next generation experiments
IceCube and AUGER, can test the viability of top-down scenarios which
generate the highest energy cosmic rays. If a signal is found soon,
future high statistics experiments should be able to map out the
neutrino spectrum, thereby allowing us direct experimental access to
physics at energy scales many orders of magnitude beyond the scope of
any conceivable particle collider on Earth.

\begin{acknowledgments}
We would like to thank Michael Kachelriess, John Learned and Tom
Weiler for valuable discussions.  This work was supported in part by a
DOE grant No. DE-FG02-95ER40896 and in part by the Wisconsin Alumni
Research Foundation. The work of M.D. was partially supported by the
SFB375 of the Deutsche Forschungsgemeinschaft.
\end{acknowledgments}


\begin{thebibliography}{99}
\bibitem{footnote}
Electrons quickly lose their
energy through synchrotron radiation, and therefore do not contribute
to the observed UHECR flux.
\end{thebibliography}
\end{document}